\documentstyle[prl,aps,multicol,epsf]{revtex}

\tighten 
\input epsf 

\begin{document}
\title{Properties of superconductor - Luttinger liquid hybrid systems }
\author{Rosario Fazio$^{(1,5)}$, F.W.J. Hekking$^{(2)}$, A.A. Odintsov$^{(3)}$, and
R.Raimondi$^{(4,5)}$}
\address{$^{(1)}$ Istituto di Fisica, Universit\`a di Catania,
viale A. Doria 6, 95129 Catania, Italy\\
$^{(2)}$Theoretische Physik III, Ruhr-Universit\"at Bochum, 
44780 Bochum, Germany\\
$^{(3)}$Faculteit der Technische Natuurkunde, TU Delft, 
2628 CJ Delft, The Netherlands\\
$^{(4)}$Dip. di Fisica "E. Amaldi", Universit\`a di Roma3,
Via della Vasca Navale 84, 00146 Roma, Italy\\
$^{(5)}$Istituto Nazionale di Fisica della Materia (INFM),Italy \\
}
\date{\today}
\maketitle

\begin{abstract}
In this paper we review some recent results concerning the physics of
superconductor - Luttinger liquid proximity systems. We discuss both
equilibrium (the pair amplitude, Josephson current, and the local density of
states) and nonequilibrium (the subgap current) properties.
\end{abstract}


\begin{multicols}{2}

\section{Introduction}

\label{intro} The properties of a normal metal (N) which is in good electric
contact with a superconductor (S) are strongly modified, a phenomenon known
as the proximity effect~\cite{deutscher65}. This effect is due to the
presence in N of Cooper pairs leaking from S, thereby giving rise to a
nonvanishing local pair amplitude. The microscopic origin for charge
transfer across the N-S interface is the phenomenon of Andreev
reflection~\cite{andreev64}. The distance over which the presence of the
superconductor
is felt in N is determined by the length $\xi _N$ over which the two
electrons forming the pair remain correlated. This length decreases with
increasing temperature $T$. In particular, for clean normal metals, $\xi
_{N} = v_F/T$, where $v_F$ is the Fermi velocity. The proximity effect
manifests itself in various ways. An example is the Josephson effect,
occurring in S-N-S sandwiches as long as the thickness of the N-layer does
not exceed $\xi _N$~\cite{aslamazov68}. Another example is the local single
particle density of states (DOS) of the metal, which acquires an energy
dependence similar to the well-known BCS-DOS of a superconductor up to
distances of the order of $\xi _N$ away from the N-S interface~\cite
{golubov88,foot1}.

Due to the recent development of superconductor-semiconductor (S-Sc)
integration technology, a revived interest arose in the properties of clean
proximity systems. Present-day high-quality Sc heterostructures combine a
number of attractive low-temperature electronic properties. The elastic mean
free path can be as long as 20$\mu$m. In typical low-density systems, the
Fermi wavelength $\lambda _F$ is of the order of 50nm. These lengths are
much larger than the corresponding ones in an ordinary metal. In addition,
the dephasing length $L_\phi$, over which the phase coherence of a single
electron is maintained, can easily be of the order of 40$\mu$m. Thus,
quantum ballistic electron propagation dominates in small-scale Sc systems.
The interplay between phase-coherent electron propagation in Sc and
macroscopic phase coherence in S gives rise to interesting new physics~\cite
{hekking94,beenakker95,lambert97,volkov97}.

During the past years, an increasing interest developed in the effects of
electron-electron interactions on the properties of low-density Sc
nanostructures. The key point is that the Coulomb energy can become
comparable to the kinetic energy, i.e., the parameter 
$e^{2}/v_{F} \sim \lambda _{F}/ a_{B}$,
where $e$ is the electron charge and $a_{B}$ is the Bohr radius, 
is no longer small. Moreover, screening
becomes less effective as the system dimensionality is decreased. As a
result a non-perturbative, microscopic treatment of interactions is
required. For one-dimensional (1D) systems this can be done in the framework
of the Luttinger model~\cite{voit95}. In a 1D interacting electron system,
also referred to as a Luttinger liquid (LL), there are no fermionic
quasiparticle excitations. Instead, the low energy excitations of the system
consist of independent long-wavelength oscillations of the charge and spin
density which propagate with different velocities. The properties of such a
system therefore are strikingly different from those of a non-interacting 1D
system. The Luttinger model is believed to be relevant for a description of
transport in a number of physical systems like Sc quantum wires and edge
states in the quantum Hall effect~\cite{fisher97}. We should also mention
the recent rapid advances in controlled fabrication of single-wall 
nanotubes~\cite{thess96}. Luttinger liquid behavior is expected in metallic
nanotubes with 
two gapless 1D modes of excitations \cite{egger97}.

At present, relatively little is known about the influence of
electron-electron interactions on the proximity effect in clean mesoscopic
N-S systems. In view of the above one may conclude that S-Sc
heterostructures are good candidates to study such effects. Of particular
interest are 1D quantum wires, connected to a superconductor (S-LL systems).
Various techniques are available nowadays to confine electrons in a
semiconductor to a long and narrow channel. In combination with
state-of-the-art S-Sc integration technology, this will make systematic
studies of interaction effects in S-LL systems feasible in the near future.
In this paper we review various properties of S-LL proximity systems. The
paper is organized as follows: In Section~\ref{luttinger} we define the
spin-1/2 Luttinger model. In Section~\ref{S-LL}, we discuss two possible
ways to couple the LL to a superconductor: via a highly transmissive and via
a tunneling interface. The remainder of the review will be divided into two
parts.
In Sections~\ref{proxi}, \ref{Josef}, and~\ref{DOSsec} we discuss equilibrium
properties of S-LL heterostructures: the pair amplitude, the Josephson effect, 
and the local density of states. In Sections~\ref{andreev} and~\ref{rgsec} we
will focus on transport phenomena and calculate the subgap current 
through a S-LL interface.


\section{The spin-1/2 Luttinger liquid}

\label{luttinger} 
In this Section we set the notation which is needed in the rest of the
paper. For a detailed discussion of Luttinger liquids we refer to existing
reviews on the topic~\cite{voit95}. The long-wavelength Hamiltonian of a 1D
interacting electron system of length $L$ can be expressed as that of a
harmonic fluid for the charge ($j=\rho$) and spin ($j=\sigma$) degrees of
freedom 
\begin{equation}
\hat{H}_L = \int \limits \frac{dx}{\pi } \sum _{j= \rho ,\sigma} v_j \left[ 
\frac{g_j}{2} (\nabla \phi _j )^2 + \frac{2}{g_j} (\nabla \theta _j)^2
\right] .  \label{lutham}
\end{equation}
The parameters $g_j$ are related to the interaction strength ($g_j =2$ for
non-interacting electrons), and $v_j=2v_F/g_j$ are the velocities of spin
and charge excitations. The commutation relation between the Bose fields 
$\theta(x)$ and $\phi(x)$ for each spin sector (spin up $s=+1$ and down 
$s=-1$) is $[\phi _s (x), \theta _{s^{\prime}} (x^{\prime})] = (i \pi /2) 
\mbox{
sign} (x^{\prime}-x) \delta_{s,s^{\prime}} , $ where $\phi _{s}= \phi
_{\rho} +s \phi _{\sigma}, $ and $\theta _{s}= \theta _{\rho} +s
\theta_{\sigma}. $

The parameters $g_j$ can be determined once one defines an appropriate
microscopic Hamiltonian. For a Sc quantum wire with spin-independent
interactions, one may take $g_\sigma =2$ and 
$g_\rho = 2/\sqrt{1+2V_0/\pi v_F}$, where $V_0$ is the zero-momentum Fourier
component of the interaction
potential~\cite{intnote}. Spin and charge excitations propagate with
different velocities ({\em spin-charge separation}). In general there will
be additional nonlinear terms appearing in Eq.~(\ref{lutham}) due to
backscattering or Umklapp processes. Usually, in quantum wires away from
half filling and in the ballistic regime these terms can be ignored;
therefore we will not discus them in the following. The electron field
operator $\hat{\Psi}$ is expressed in terms of the boson field in terms of
the spin and charge degrees of freedom 
\begin{equation}
\hat{\Psi} _{s}(x) \sim \sqrt{\rho _0} \sum \limits _{\delta = \pm} e^{i
\delta k_F x } e^{i \delta [\theta _{s} -\delta \phi _{s}]}
\label{fieldoperator}
\end{equation}
where $k_F$ is the Fermi wave vector and $\rho_0 = k_F/\pi$ the
electron density.


\section{Superconductor-Luttinger liquid interfaces}

\label{S-LL} 

A superconductor can contact a quantum wire at the end (edge contact) or at
some internal point or segment of the wire (lateral contact). Both edge and
lateral contacting have been already implemented in experiments with S-Sc
structures~\cite{Nitta95}. The quality of S-LL interfaces can be
characterized by the transparency of the barrier at the interface. We will
consider below edge contacts with both high and poor transparency of the
barriers as well as poorly transmitting lateral point contacts.

{\underline{Perfectly transmitting interfaces} - Maslov et {\em al.}~\cite
{maslov96} and Takane and Koyama~\cite{takane96} recently developed a
bosonization scheme to treat clean S-LL interfaces. Following Ref.~\cite
{maslov96} we consider the case of two superconductors, kept at a phase
difference $\chi$, and adiabatically connected to a quantum wire of length 
$L $ (the results for a single interface can be obtained by taking the limit 
$L \to \infty$). The bosonization scheme can be carried out once the boundary
conditions for the right and left moving fields at the S-LL interface are
determined. }

The mode expansions for the fields $\theta_j$ and $\phi_j$ ($j = \rho,
\sigma $), such that the Fermi operators (\ref{fieldoperator}) satisfy the
proper boundary conditions at the interface, are 
\begin{eqnarray}
\theta _{\rho}(x) & = & \varphi_\rho + \sqrt{\frac{g_{\rho}}{2}} \sum _{q >
0} \gamma_q^+ (\hat{b}^{\dagger}_{\rho,q} + \hat{b}_{\rho,q}) ,
\label{phitheta1} \\
\theta _{\sigma}(x) & = & \frac{\pi x}{4 L} 
M_\sigma + \sqrt{\frac{g_{\sigma}}{2}} \sum _{q > 0} \gamma_q^- (\hat{b}^{\dagger}_{\sigma,q} - \hat{b}_{\sigma,q}) ,  \label{phitheta2} \\
\phi_{\rho}(x) & = & \frac{\pi x}{4 L} (1+\frac{J_\rho}{2} + \frac{\chi}{\pi}) + \sqrt{\frac{2}{g_{\rho}}} \sum _{q > 0} \gamma_q^- (\hat{b}
^{\dagger}_{\rho,q} - \hat{b}_{\rho,q}) ,  \label{phitheta3} \\
\phi _{\sigma}(x) & = & \varphi_{\sigma} + \sqrt{\frac{2}{g_{\sigma}}} \sum
_{q > 0} \gamma_q^+ (\hat{b}^{\dagger}_{\sigma,q} + \hat{b}_{\sigma,q}) ,
\label{phitheta4}
\end{eqnarray}
where $q = \pi n/L$ $\gamma_q^+ = (\pi/2qL)^{1/2}\cos(qx)$, and $\gamma_q^-
= i(\pi/2qL)^{1/2}\sin(qx)$. The system is characterized by the two
topological numbers $J_\rho$ and $M_\sigma$; the topological numbers $N_s
\equiv (M_\sigma \pm J_\rho)/2$ must be odd integers \cite{maslov96}.

Substituting the mode expansions (\ref{phitheta1}) -- (\ref{phitheta4}) into
the phase Hamiltonian (\ref{lutham}), we obtain a Hamiltonian of the form 
\begin{eqnarray}
H & = & \frac{\pi}{4L} \left[ \frac{ g_{\rho}v_{\rho} }{2} \left( 
\frac{J_{\rho}}{2} + \frac{\chi}{\pi} +1 \right)^2 +
\frac{2v_{\sigma}}{g_{\sigma}} 
\left( \frac{M_{\sigma}}{2} \right)^2 \right]  \nonumber \\
& + & \sum_{j = \rho, \sigma} \sum_{q>0} v_j q \hat{b}^{\dagger}_{j,q} 
\hat{b}_{j,q}.  \label{H_top}
\end{eqnarray}
Takane and Koyama~\cite{takane96} extended this bosonization scheme to
include the energy dependence of the phase shift related to Andreev
reflection, thereby showing that it may be important if a sufficiently
strong potential barrier is placed at the S-LL interface. In the following
we will neglect this energy dependence.

\underline{Poorly transmitting interfaces} - If the quantum wire is weakly
coupled to a superconductor at $x=0$, the system can be treated in terms of
the tunnel Hamiltonian formalism. The Hamiltonian of the whole 
system\cite{fazio95&96}, $\hat{H} = \hat{H}_{S} + \hat{H}_L + \hat{H}_T $,
contains the BCS-Hamiltonian $\hat{H}_S$ describing the bulk superconductor,
the Hamiltonian $\hat{H}_L$ of the quantum wire, and the tunnel Hamiltonian, 
\begin{equation}
\displaystyle{\hat{H}_T = \sum _{s} t_0 \hat{\Psi} ^{\dagger}_{S,s} (x=0) 
\hat{\Psi}_{L,s}(x=0)} + \mbox{(h.c.)}.  \label{tunnelham}
\end{equation}
The tunnel matrix elements $t_0 $ can be related to the tunnel conductances 
$G_T$ of the junctions, $G_T = 4\pi e^2 N_L(0) N_S (0) |t_0|^2 $, where 
$N_L(0)= 1/\pi v_F$, and $N_S(0)$ is the normal
state density of states at the Fermi level of the superconductor.

If two superconductors are attached to the ends of a quantum wire one should
use the recently developed bosonization technique for finite 1D systems with
open boundaries~\cite{fabrizio95}. The presence of lateral tunnel contacts
does not impose any additional boundary conditions on the Fermi fields. We
will neglect possible inhomogeneities of the quantum wire near the contacts,
which might be the source of electron backscattering. For a discussion of
possible modifications due to the presence of a barrier potential, see 
Ref.~\cite{takane97}.

\section{Pair amplitude}

\label{proxi}

In this section we will evaluate the pair amplitude induced into a Luttinger
liquid which is connected to a superconductor at $x=0$. We calculate the
pair amplitude at a distance $x$ from the contact. The pair amplitude is
defined as the anomalous time-ordered expectation value $\Xi (x,\tau) \equiv
- \langle T_{\tau}\psi_{\uparrow} (x, \tau ^+)\psi_{\downarrow} (x, \tau) \rangle $, 
where $\tau^{+} $ tends to $\tau$ from above. Throughout this section, we will be
interested in the smooth spatial variation of $\Xi$ (we ignore spatial
variations on the scale of $\lambda_F$). We will evaluate $\Xi$ both for a
highly transmissive and for a tunneling interface.

\underline{Perfectly transmitting interfaces} - The pair amplitude for an
adiabatic interface between S and LL has been obtained by Maslov et al.~\cite
{maslov96}: $\Xi (x) = - 2 \rho _{0} \langle \exp{2i \phi _{\rho}}\rangle
\langle \cos 2\theta _{\sigma}\rangle . $ Since Maslov's bosonization
procedure has been developed for Andreev scattering at energies much smaller
than the superconducting gap $\Delta$, we shall evaluate $\Xi$ at large
distances from the interface, i.e.,
with a short wavelength cut-off $\alpha$ such that 
$x \gg ·\alpha \sim \xi _{S} \equiv v_{F} /
\Delta$. Performing the averages with respect to (\ref{H_top}), we obtain 
\begin{equation}
\Xi (x) = - 2 \rho _{0} \left[\frac{\pi \alpha /\beta v_{F}}{\sinh 2\pi x
/\beta v_{F}}\right]^{1/2} \left[\frac{\pi \alpha /\beta v_{\rho}}{\sinh
2\pi x /\beta v_{\rho}}\right]^{1/g_{\rho}} ,
\end{equation}
where $\beta =1/T$. The result consists of a product of a spin (first term
in brackets) and a charge contribution (last term in brackets); only the
latter is sensitive to interactions.

At zero temperature, the pair amplitude for non-interacting electrons decays
slowly away from the junction for: $\Xi _{T=0} (x) \sim 1/x$. Nothing
prevents the superconducting correlations from being present arbitrarily
deep in the quantum wire. In the presence of repulsive interactions, the two
electrons can be scattered out of their time-reversed state and therefore 
$\Xi$ decays faster with increasing distance $x$ from the interface, 
\begin{equation}
\Xi _{T=0} \sim \rho _{0} \left[\frac{\alpha}{2x}\right]^{1/2+1/g_{\rho}} .
\end{equation}

At finite temperatures, the coherence length in the LL becomes finite, and,
correspondingly, the pair amplitude should decay faster. This can be seen by
calculating the ratio $\Xi _{T}/\Xi _{T=0}$. At low temperatures $T \ll
v_{F}/x$, this ratio behaves as: 
\begin{equation}
\frac{\Xi _{T}(x)}{\Xi _{T=0}(x)} \simeq 1 - \frac{1}{6} 
\left(\frac{\pi x T}{v_{F}}\right)^{2},
\end{equation}
independent of the interaction strength. At higher temperatures $T \gg
v_{F}/x$, the suppression will be exponential: 
\begin{equation}
\frac{\Xi _{T}(x)}{\Xi _{T=0}(x)} \simeq
\left(\frac{g_{\rho}}{2}\right)^{1/g_{\rho}} \left(\frac{2\pi \alpha
    T}{v_{F}}\right)^{\frac{1}{2}
+ \frac{1}{g_{\rho}}} e^{(-2\pi x T /v_{F})}.
\end{equation}
Here, interactions determine the pre-exponential temperature dependence.

\underline{Poorly transmitting interfaces} - In order to evaluate the pair
amplitude for a tunnel interface between S and LL, we use standard
perturbation theory in the tunnel Hamiltonian $H_{T}$, Eq.~(\ref{tunnelham}). 
In the tunneling regime the usual bosonization scheme can be used, with
short wavelength cut-off $\alpha \sim \lambda _{F}$. The lowest order
non-vanishing contribution to the pair potential in the LL is given by the
following expression 
\begin{eqnarray}
\Xi(x,\tau) = t_0^{2} \int \limits _{0}^{\beta} d\tau _{1} & d\tau _{2} &
\Pi _{+,-,-,+}(x,\tau ^{+};x,\tau; 0,\tau _{1};0, \tau _{2})  \nonumber \\
& \times & F_{+,-}(0,\tau_{2};0,\tau_{1}) .  \label{Xi}
\end{eqnarray}
Here $F_{+,-}(0,\tau _2;0,\tau _1)$ is the usual anomalous Green's function
for the superconductor, and we introduced the four-point correlator $
\Pi_{1,2,3,4}(1,2,3,4) \equiv \langle T_{\tau} \psi _{s1}(1)\psi _{s2}(2)
\psi ^{\dagger}_{s3}(3) \psi ^{\dagger}_{s4}(4) \rangle. $ The relevant
process consists in the tunneling of two electrons from S into the LL. This
process is of second order in the tunneling. First, an electron tunnels into
the LL, leaving behind a quasiparticle excitation in S. Then, the second
electron tunnels from S into LL, annihilating the quasiparticle in S. Thus,
two time scales play a role in this tunneling process: the lifetime $|\tau
_{1} - \tau _{2}| \sim 1/\Delta$ of the intermediate state with one
quasiparticle in S, and the time $|\tau - \tau _{1}|$, $|\tau - \tau _{2}|
\sim |x|/v_{F}$ during which each electron propagates in the LL to the
position $x$. The behavior of the pair amplitude will depend on the relative
magnitude $|x|\Delta /v_{F}$ of these time scales for propagation and
tunneling.

\underline{Pair amplitude close to the junction} - For distances close to
the junction, $\xi _{S}\gg |x|\gg \alpha \sim \lambda _{F}$, the time needed
for the pair to traverse the LL is negligible compared to the quasiparticle
lifetime $1/\Delta $. Hence, the main contribution to $\Xi $ stems from
electrons that propagate fast and independently through the LL. In this
case, we can approximate the four-point correlator by the product of two
Green's functions $\Pi \simeq G_{-,-}(x,\tau -\tau _{1})G_{+,+}(x,\tau
^{+}-\tau _{2})$ with $G_{s,s^{\prime }}(x,\tau )\simeq \delta _{s,s^{\prime
}}G(x)\delta (\tau ).$ The function $G(x)$ is obtained by integrating the
single particle Green's function $G_{s,s^{\prime }}(x,\tau )$ over imaginary
time (see Appendix A). Within this approximation we find $\Xi (x,\tau
)\simeq t_{0}^{2}G^{2}(x)F_{+,-}(0,\tau ^{+};0,\tau ).$ The anomalous
Green's function can be written as $F_{+,-}(0,\tau ^{+};0,\tau
)=N_{S}(0)\Delta \sinh ^{-1}(\omega _{D}/\Delta ),$ where the
Debye-frequency $\omega _{D}$ is a high energy cut-off in S. We thus arrive
at 
\begin{equation}
\Xi (x)=\Xi ^{0}\left[ \frac{g_{\rho }}{2}\right] ^{1/g_{\rho }+g_{\rho
}/4}\left[ \frac{\pi \alpha }{\beta v_{F}}\right] ^{1/g_{\rho }+g_{\rho
}/4-1}F_{G}^{2}(x,g_{\rho }),
\end{equation}
where 
\begin{equation}
\Xi ^{0}\equiv 2\rho _{0}\frac{G_{T}}{(4e^{2}/\pi )}\frac{\alpha \Delta }
{v_{F}}\sinh ^{-1}(\omega _{D}/\Delta ).
\end{equation}
Expressions for the function $F_{G}(x,g_{\rho })$ are given in Appendix A.

In the noninteracting case $g_{\rho }=2$ the pair amplitude becomes
space-independent at zero temperature, $\Xi _{g_{\rho }=2}(x)\to \Xi ^{0}$.
This reflects the fact the time-reversed electrons from S reach points in
the LL within a distance $\sim \xi _{S}$ from the junction instantaneously.
In the presence of repulsive interactions, propagation through the LL
becomes faster, $v_{\rho }>v_{F}$, and one naively would expect the pair
amplitude to remain space independent. However, for $g_{\rho }<2$ we find at
zero temperature, 
\begin{equation}
\Xi _{T=0}(x)=\Xi ^{0}\left[ \frac{\alpha }{x}\right] ^{1/g_{\rho }+g_{\rho
}/4-1}F_{G}^{2}(g_{\rho }).  \label{pairamp_close}
\end{equation}
Due to repulsive interactions, the pair amplitude becomes $x-$dependent: it
decays algebraically away from the junction. This suppression of the pair
amplitude is related to the fact that the effective tunneling amplitude of
Cooper pairs is renormalized at low energies by the interactions, similar to
the tunneling amplitude of single electrons into a LL~\cite{fisher97}.

At finite temperatures, we therefore expect a competition between two
effects: (i) enhancement of the tunneling due to thermal fluctuations
leading to an {\em enhancement} of the pair amplitude as a function of
temperature; (ii) decreasing coherence length, leading to a {\em suppression}
of the pair amplitude with temperature. As a result of the above
competition, the pair amplitude shows a maximum as a function of temperature
if $g_{\rho} <2$. Analytical results can be obtained in the case of weak
interactions $(1-g_{\rho}/2)\pi x/\beta v_{F} \ll 1$. At low temperatures 
$\beta v_{F} \gg x$ 
\begin{equation}
\frac{\Xi _{T}}{\Xi _{T=0}} \simeq 1+ (2-g_{\rho}) \frac{x}{\beta v_{F}} -
\frac{2}{\sqrt{\pi}}\frac{\Gamma(\nu +1/2)} {\Gamma(\nu +1)}\left(\frac{x}
{\beta v_{F}}\right)^{2\nu}
\end{equation}
with $\nu = 1/4g_{\rho} + g_{\rho}/16 +1/4$. Note that the first term,
describing the thermal enhancement of the pair amplitude, vanishes in the
noninteracting case $g_{\rho}=2$, as it should.

\underline{Pair amplitude away from the junction} - If we are interested in
the pair amplitude at large distances from the interface, $|x| \gg \xi _{S}$, 
we can neglect the time $1/\Delta$ spent in the virtual state in
comparison with the long time $x/v_{F}$ needed to traverse the LL. We thus
approximate the anomalous Green's function in S with the help of a 
$\delta$-function in time as 
\begin{equation}
F_{+,-}(0,\tau_{1};0,\tau_{2}) \approx \pi N_S(0) \delta (\tau _{1} -
\tau_{2}).  \label{deltaaprox}
\end{equation}
This enables us to perform one integration over imaginary time in Eq.~(\ref
{Xi}), and obtain 
\begin{equation}
\Xi(x) = \pi N_S(0) t_0^{2} \int \limits _{0}^{\beta} d\tau \Pi _C(x,\tau).
\end{equation}
The function $\Pi _{C}(x,\tau)$ is defined in Appendix B.

In the noninteracting case $g_{\rho} = 2$, the propagator $\Pi _{C}$ can be
integrated analytically. In the zero temperature limit $\beta \to \infty$,
the pair amplitude is given by 
\begin{equation}
\Xi _{g_{\rho} =2} (x) = 2 \rho _{0} \frac{G_{T}}{(4e^{2}/\pi)} \frac{\alpha}
{x} .  \label{XinonzeroT}
\end{equation}
We see that $\Xi$ decays slowly away from the junction. With increasing
temperature, the distance $v_{F}/T$ over which two electrons maintain their
relative phase coherence decreases, and in the limit $\beta v_{F}/|x| \ll 1$
the pair amplitude decays exponentially with $x$: 
\begin{equation}
\Xi _{g_{\rho =2}} (x) \simeq 4 \rho _{0} \frac{G_{T}}{(4e^{2}/\pi)} \frac{2
\pi \alpha}{\beta v_{F}} \exp{(-2\pi x/\beta v_{F})} .  \label{XinonhiT}
\end{equation}

We now turn to the case of repulsively interacting electrons, $g_{\rho }<2$.
In the zero temperature limit we obtain 
\begin{equation}
\Xi (x)=2\rho _{0}\frac{G_{T}}{(4e^{2}/\pi )}\left[ \frac{\alpha }{x}\right]
^{2/g_{\rho }}F_{C}(g_{\rho }),  \label{pairamp_far}
\end{equation}
where $F_{C}(g_{\rho })$ is given in Appendix B. We find a power law decay
of $\Xi $ with $x$; compared to the noninteracting case, Eq.~(\ref
{XinonzeroT}), the pair amplitude decays faster away from the junction. In
the high temperature limit $\beta v_{F}/|x|\ll 1$, the decay becomes
exponential, 
\begin{equation}
\Xi (x)\simeq 4\rho _{0}\frac{G_{T}}{(4e^{2}/\pi )}\left[ \frac{2\pi \alpha 
}{\beta v_{\rho }}\right] ^{2/g_{\rho }}\exp {(-2\pi x/\beta v_{F})}.
\label{pairamphighT}
\end{equation}
Note, however, the algebraic temperature dependence of $\Xi $ through the
pre-\-ex\-po\-nen\-tial factor.

\section{Josephson effect}
\label{Josef} 

The Josephson effect is an observable consequence of the penetration
of the pair amplitude into LL \cite{fazio95&96,maslov96,takane97}. The
calculation of the Josephson current through S-LL-S system is analogous to
the one for the pair amplitude (see Section \ref{proxi}). For perfectly
transmitting interfaces \cite{maslov96} the critical current is equal to its
value for non-interacting electrons independent of the actual interaction in
LL. This result stems from the fact that features of LL are observable in
transport experiments {\em only} in the presence of backscattering, both in
normal \cite{fisher97} and superconducting systems.

The case of poorly transmitting interfaces has been treated in Refs. 
\cite{fazio95&96} within the tunnel Hamiltonian formalism. An infinite
quantum wire coupled to two superconductors by tunnel junctions at a
distance $x$ was considered. In the lowest order in $\hat{H}_T$
(\ref{tunnelham}), the DC Josephson current $I_{J}=I_{c}\sin \chi $ shows
standard dependence on the phase difference $\chi $ between the
superconductors. The
critical current $I_{c}$ can be estimated as,
\begin{equation}
I_{c}\sim \frac{G_{T1}G_{T2}}{(2e^{2}/\pi )^{2}}e\delta \Im _{c},
\label{Icest}
\end{equation}
where the characteristic energy scale $\delta $ and the scaling factor $\Im
_{c}$ characterizing the decay of the Cooper pair density in the LL are
given by $\delta =\Delta $, $\Im _{c}=(\alpha /x)^{g_{\rho }/4+1/g_{\rho
}-1}$, for $x\ll \xi _{S}$, and $\delta =v_{F}/x$, $\Im _{c}=(\alpha
/x)^{2/g_{\rho }-1}$, for $x\gg \xi _{S}$ (cf. Eqs. (\ref{pairamp_close}), 
(\ref{pairamp_far})). Similarly to the pair amplitude, the critical current
shows maximum as a function of temperature for both $x\ll \xi _{S}$ and 
$x\gg \xi _{S}$. At high temperatures $T\gg v_{F}/x$ the Josephson current is
suppressed exponentially (cf. Eq. (\ref{pairamphighT})). Numerical estimates 
\cite{fazio95&96} show that for typical experimental parameters the critical
current (\ref{Icest}) is in a few nanoamp range.

If a DC voltage $eV\ll 2\Delta $ is applied between the superconductors, the
Josephson phase difference $\chi $ becomes time-dependent, 
$\dot{\chi} =\omega _{J}=2eV$, and the Josephson current oscillates with the frequency $\omega _{J}$ (AC-Josephson effect). For non-interacting electrons the
Josephson current $I_{J}(t)\propto I_{ac}\sin \left( \omega _{J}t-\chi
_{0}\right) $ acquires additional phase shift $\chi _{0}=eVx/v_{F}$ due to
the propagation of electrons between the contacts. In the interacting case
also the amplitude $I_{ac}$ of AC current becomes voltage dependent. Namely,
for moderate interaction the amplitude $I_{ac}(V)$ shows pronounced
oscillations with the period $eV_{0}=2\pi /[x/v_{\rho }-x/v_{\sigma }]$
corresponding to $2\pi $ difference between the phases of charge $\rho $ and
spin $\sigma $ excitations \cite{fazio95&96}. Therefore, the AC Josephson
effect can be used as a tool for the observation of the {\em spin-charge
separation} in LL.

A different geometry - finite quantum wire of length $x\gg \xi _{S}$
contacting the superconductors at the ends - was considered in Refs. \cite
{maslov96},\cite{takane97}. The result by Maslov et. al. \cite{maslov96} ($\Im
_{c}=(\alpha /x)^{2/g_{\rho }-1}$) was revised by Takane \cite{takane97},
who found stronger decay of the critical current, $\Im _{c}=(\alpha
/x)^{2(2/g_{\rho }-1)}$, due to a pinning of spin fluctuations by
superconductors.

\section{Density of states}
\label{DOSsec} 
The local DOS is defined through the retarded one-electron Green's function
of the LL $G_R (x,x^{\prime};t) \equiv -i\langle
\{\psi(x,t),\psi(x^{\prime},0) ^{\dagger}\}\rangle \theta (t) , $ as 
\begin{equation}
N(x,\omega ) = - \frac{1}{\pi} {\cal I} \mbox{m} \int _{-\infty} ^{\infty}
dt e^{i\omega t} G_R(x,x;t) \:\:\:\:\:\: .  \label{DOS}
\end{equation}
In the case of an infinite LL it is straightforward to compute this quantity
and get~\cite{voit95} $N(\omega ) \sim \omega ^{(g_{\rho}+4/g_{\rho}-4)/8} . 
$ Contrary to Fermi liquids, whose quasiparticle residue is finite,
Luttinger liquids have a density of states which vanishes at the Fermi
energy as a power law, both for repulsive ($g_{\rho} < 2$) and attractive 
($g_{\rho} > 2$) interactions. In the non-interacting case ($g_{\rho} = 2$)
the density of states is constant, as for a Fermi liquid~\cite
{voit95,fisher97}.

What are the modifications of the local DOS due to proximity effect? Here we
discuss the space and frequency dependence of the DOS of a LL contacted at 
$x=0$ with a superconductor~\cite{winkelholz96}, which corresponds to the
limit $L \to \infty$ in the mode expansion given by Eqs.~(\ref{phitheta1})
-- (\ref{phitheta4}). In this case only the non-zero modes contribute to the
local DOS. The correlation function $\langle
\psi^{\dagger}(x,t)\psi(x,0)\rangle$ can be evaluated using the boson
representation (the correlator is not translationally invariant due to the
presence of the interface). At small energies the DOS behaves as 
\begin{equation}
N_{S-LL}(\omega ) \sim \omega^{g_{\rho}/4-1/2} .  \label{LLSDOS}
\end{equation}
The exponent of the DOS is negative ($g_{\rho} < 2$) and, hence, there is a 
{\em strong enhancement} at low energies whereas in the absence of S the LL
would show a {\em vanishing} DOS at the Fermi energy. The enhancement of the
DOS occurs {\em regardless} of the distance $x$ from the interface. The
scale of the enhacement is set by a space dependent high frequency cutoff 
$\omega _c \sim v_{\rho}/x$. The induced pair amplitude in the LL, which is
characteristic of the presence of the superconductor, decays as a power~\cite
{maslov96} of the distance $x$ (see Section~\ref{proxi}). This profound
difference in the space dependence demonstrates that the DOS provides
different information compared to the proximity effect. The reason why the
DOS does not approach the well-known behaviour of an Luttinger liquid far
from the superconducting contact is in part related to the fact that we are
considering a clean wire. In this case the states in the LL are extended and
the DOS enhancement does not depend on $x$.

So far we discussed only the case in which the interface between the
superconductor and the Luttinger liquid has a high transparency. Let us
shortly comment on the opposite limit, in which the Luttinger liquid is
connected to the superconductor by a tunnel junction. At low energies, we
find for the DOS close to the junction $N_{S-LL} \sim \omega ^{(g_{\rho}/2
-1) + (1/2g_{\rho} - g_{\rho}/8)} $. Although the exponent is different from
the one appearing in Eq.~(\ref{LLSDOS}), the DOS is clearly enhanced.
Moreover, also in this case the enhancement is found regardless of the
distance from the junction.


\section{Two-electron tunneling into a Luttinger liquid}

\label{andreev} 
In this Section, we will calculate the subgap conductance for a LL,
connected to S via a lateral tunnel junction~\cite{fazio95b}. The tunnel
current is calculated in the standard way as 
\begin{equation}
I(t) = -e\langle\dot{N}_{L}(t)\rangle = -ie \langle [H_T(t),N_L(t)]\rangle .
\end{equation}
Here, $N_L = \sum _s \int dx \psi _{L,s}^{\dagger}(x) \psi _{L,s}(x)$; the
time-dependent tunnel Hamiltonian is given by 
\begin{equation}
H_T(t) = \sum \limits _{s} e^{ieVt} t_0 \psi _{L,s} ^{\dagger}(0,t) \psi
_{S,s} (0,t) + \mbox{h.c.},  \label{timetunnel}
\end{equation}
where $eV$ is the applied bias voltage between S and LL. The time dependent
field operators are defined as 
\[
\psi _{i,s} (x,t) = e^{i(H_i - \mu _iN_i)t} \psi _{i,s} (x) e^{-i(H_i - \mu
_i N_i)t}, 
\]
where $\mu _i$ is the chemical potential. We then perform an expansion in
the tunneling Hamiltonian (which is switched on adiabatically at $t=-\infty$),
and obtain 
\begin{equation}
I(t) = -ie \int \limits _{-\infty}^{\infty} dt_1 dt_2 dt_3
G_R(t,t_1,t_2,t_3) ,
\end{equation}
where we introduced the retarded Green's function 
\begin{eqnarray}
G_R(t,t_1,t_2,t_3) = i\theta(t-t_1) \theta(t_1-t_2) \theta(t_2-t_3) 
\nonumber \\
\times \langle [[[[H_T(t),N_L(t)],H_T(t_1)],H_T(t_2)],H_T(t_3)] \rangle .
\end{eqnarray}

Using imaginary time techniques, $G_R$ can be expressed as a product of
time-ordered correlation functions for LL and S.
As long as the relevant energies ($eV$, $T$) are small compared to the gap
of the superconductor, the time-dependence of the anomalous correlations in
the superconductor can be approximated with the help of $\delta$-functions,
see Eq.~(\ref{deltaaprox}). In the zero temperature limit we get 
\begin{eqnarray}
I = 24 e^2 (\rho_0 \alpha)^2 V \frac{G_T^2}{(4e^2/\pi)^2} \left[
  \frac{(g_\rho/2)^{\alpha_1} (\alpha eV/v_F)^{\alpha_1-2}} {\Gamma(\alpha_1)}
\right.  \nonumber \\
\left. + \frac{(g_\rho/2)^{\alpha_2}(\alpha eV/v_F)^{\alpha_2-1}} 
{\Gamma(\alpha_2+1)} \right] ,
\end{eqnarray}
where $\alpha_1 = 2/g_\rho+g_\rho /2$ and $\alpha_2 =2/g_\rho $ Note that
the current depends on the applied bias in a power law fashion, which is
common for transport through an interacting 1D system~\cite{fisher97}. For
noninteracting electrons $I \sim V$ as one expects. For repulsive
interactions $g_\rho <2$, and the dominant contribution to current at low
voltages reads $I \sim V^{2/g_\rho}$.

At finite temperatures we have 
\begin{equation}
G= \frac{1}{4}G_{T}^{2}R_{K}\left[c_{1}
  \left(\frac{T}{E_{F}}\right)^{\frac{g_{\rho}}{2}  +\frac{2}{g_{\rho}}-2}
+c_{2}\left(\frac{T}{E_{F}}\right)^{\frac{1} {g_{\rho}}-1} \right] \; .
\end{equation}
Here, $E_F$ is the Fermi energy; $c_{1},c_{2}$ are dimensionless constants,
which can be determined numerically (some examples are $c_1=c_2=1.0$ for 
$g_{\rho}=2.0$, $c_1=1.549$, $c_2=2.467$ for $g_{\rho}=1.0$, and $c_1=1.153$
and $c_2=0.765$ for $g_{\rho}=3.0$).

In the case of a chiral LL the subgap conductance has been studied by 
Fisher~\cite{fisher94}. 

\section{Renormalization Group for the subgap conductance}

\label{rgsec} 
In this last section we discuss the subgap current by means of a poor man's
renormalization group~\cite{takane96,takane97b,roberto96} previously
developed for the case of an impurity in a Luttinger liquid\cite{fisher97}.
To be specific we consider a semi-infinite normal metal for $x<0$ and a
superconductor for $x>0$. At $x=0$ there is an insulating barrier which is
modeled by a delta-like potential $U(x)=U_0 \delta (x)$. In the absence of
electron-electron interaction in the normal metal, the scattering states,
due to the presence of the interface with the superconductor, are 
\begin{eqnarray}
&\phi_k(x)={\frac{1}{{\sqrt{2\pi}}}} \left( 
\begin{array}{c}
e^{ikx}+r_0e^{-ikx} \\ 
-ir_ae^{ik^*x}
\end{array}
\right),~~~~x<0&\cr &\phi_{-k^*}(x)={\frac{1}{{\sqrt{2\pi}}}} \left( 
\begin{array}{c}
ir_ae^{-ikx} \\ 
e^{-ik^*x}+r_0^*e^{ik^*x}
\end{array}
\right),~~~x<0.&  \label{scattering}
\end{eqnarray}
The states $\phi_k$ and $\phi_{k^*}$ correspond to incoming particles and
holes; $r_0$ and $-ir_a$ correspond to the amplitudes for normal and Andreev
scattering. For the sake of simplicity, we make the Andreev approximation:
in the absence of a potential barrier (i.e., $U_0 =0$) at the interface, one
has $r_0 =0$ and $r_a=1$. For our present case of a delta-like potential 
\begin{eqnarray}
r_0 &=&-{\frac{{2zz^*+i(z+z^*)}}{{(1+iz)(1-iz^*)+zz^*}}}, \\
r_a &=&{\frac{{1}}{{(1+iz)(1-iz^*)+zz^*}}},  \label{coefficients}
\end{eqnarray}
where $z=mU_0/k$; furthermore $k=\sqrt{2m(\mu +i \omega_n)}$, and $\omega_n
=2\pi (n+1/2)T$. We work with Matsubara frequencies. To get the proper
retarded Green function eventually $i\omega_n \rightarrow \omega +i\delta$
as usual. The effect of the interaction is analyzed first in the Born
approximation. Consider for simplicity the Hartree contribution in first
order in perturbation theory. The scattering states of 
Eq.~(\ref{scattering}) generate an oscillating electron density, which in turn
generates a 
potential $V_H (x)$. The correction to the wave function can be written in
the form 
\begin{equation}
\delta\psi_{k,\alpha}(x)= \int^{0}_{-\infty} dy G_{k,\alpha\beta} (x,y)V_H
(y) \phi_{k,\beta}(y) ,
\end{equation}
where $\alpha$ and $\beta$ take the values 1 and 2 corresponding to the
particle and hole degree of freedom, respectively. The matrix elements of
the Green's function take up the values 
\begin{equation}
\begin{array}{l}
G_{k,11}(x,y)={\frac{{m}}{{i}}}\displaystyle{\ {\frac{
{(e^{ik|x-y|}+r_0e^{-ik(x+y)}}}{{k}}}}, \\ 
G_{k,12}(x,y)={\frac{{m}}{{i}}}\displaystyle{\ {\frac{{ir_ae^{-i(kx-k^*y)}}}
{{k^*}}}}, \\ 
G_{k,21}(x,y)=-{\frac{{m}}{{i}}}\displaystyle{\ {\frac{{ir_ae^{i(k^*x-ky)}}}
{{k}}}}, \\ 
G_{k,22}(x,y)=-{\frac{{m}}{{i}}}\displaystyle{\ {\frac{
{(e^{-ik^*|x-y|}+r_0^*e^{ik^*(x+y)})}}{{k^*}}}}.
\end{array}
\end{equation}
The correction to the Andreev scattering coefficient is obtained by setting 
$\alpha =2$ and $\beta =1$, and reads 
\begin{eqnarray}
\delta \psi_{k,2}(x) = \int^{0}_{-\infty} dy &[ &
G_{k,21}(x,y)V_H(y)\phi_{k,1}(y)  \nonumber \\
& + &G_{k,22}(x,y)V_H(y)\phi_{k,2}] .
\end{eqnarray}
By inserting the expression for the Green's function 
$G_{k,\alpha,\beta}(x,y) $ and the non-interacting scattering states 
$\phi_{k,\beta}(x)$
one gets $\delta r_a =\delta r_a^{(1)} +\delta r_a^{(2)}, $ with 
\begin{eqnarray}
\delta r_a^{(1)} ={\frac{{mr_a r_0}}{{i k}}} \int^{0}_{-\infty} dy V_H(y)
e^{-2iky} , \\
\delta r_a^{(2)} =- {\frac{{mr_a r_0^*}}{{i k}}} \int^{0}_{-\infty} dy
V_H(y) e^{2ik^*y}.
\end{eqnarray}
By expanding the Hartree potential in a Fourier series, one obtains an
expression for the scattering coefficients 
\begin{eqnarray}
\delta r_a^{(1)} =-{\frac{{mr_a r_0}}{{\ k}}}
\int_{-\infty}^{\infty}{\frac{{dq}}{{2\pi}}} V(q) \delta n (q)
{\frac{{1}}{{q-2k}}}, \\ 
\delta r_a^{(2)} ={\frac{{mr_a r_0^*}}{{\ k}}}
\int_{-\infty}^{\infty}{\frac{{dq}}{{2\pi}}}  V(q) \delta n (q) {\frac{{1}}{{q+2k^*}}},
\end{eqnarray}
where 
\begin{equation}
\delta n(q) =\int^0_{-\infty} dy e^{-iqy} \delta n(x).
\end{equation}
Notice that the convergence of the integral is automatically controlled by
the imaginary parts of the momenta $k$ and $k^*$. The density $\delta n( x)$
of the electron is evaluated in terms of the non-interacting scattering
states, and can be written as 
\begin{eqnarray}
n(x) &= &n_0 +\delta n(x)  \nonumber \\
&= &-{\frac{1}{2}}\left(T\sum_{\omega_n } G_{11} (x,x) +T\sum_{\omega_n }
G_{22}(x,x)\right) \;\;.
\end{eqnarray}
For the Fourier transform $\delta n(q)$ we then get 
\begin{eqnarray}
&\delta n(q) =\lim_{\delta \rightarrow 0^+}{\frac{1}{2}}\displaystyle{\
\left( {\frac{{ir_0}}{{2\pi}}} \ln\left(
    {\frac{{2k_F+q+i\delta}}{{q+i\delta}}}\right)\right.} &\cr
&\displaystyle{\left.-{\frac{{ir_0^*}}{{2\pi}}} 
\ln\left( {\frac{{q-2k_F+i\delta}}{{q+i\delta}}}\right)\right).}&
\end{eqnarray}
Hence 
\begin{eqnarray}
&\displaystyle{\delta r_a^{(1)} =-{\frac{{imr_a r_0}}{{4\pi k}}}
\int_{-\infty}^{\infty}{\frac{{dq}}{{2\pi}}} V(q) {\frac{{1}}{{q-2k}}}}&\cr 
&\times \displaystyle{\left( {r_0} \ln\left( {\frac{{2k_F+q+i\delta}}{{
q+i\delta}}}\right) -{r_0^*} \ln\left( {\frac{{q-2k_F+i\delta}}{{q+i\delta}}}
\right)\right)},&  \nonumber
\end{eqnarray}
and 
\begin{eqnarray}
&\displaystyle{\delta r_a^{(2)} ={\frac{{imr_a r_0^*}}{{4\pi k}}}
\int_{-\infty}^{\infty}{\frac{{dq}}{{2\pi}}} V(q) {\frac{{1}}{{q+2k^*}}}}&
\cr &\times \displaystyle{\left( {r_0} \ln\left( {\frac{{2k_F+q+i\delta}}{{
q+i\delta}}}\right) -{r_0^*} \ln\left( {\frac{{q-2k_F+i\delta}}{{q+i\delta}}}
\right)\right).}&  \nonumber
\end{eqnarray}
Using the Cauchy theorem, we find 
\begin{eqnarray}
&\displaystyle{\delta r_a^{(1)} = {\frac{{mr_a r_0}}{{4\pi k}}} \left( r_0
V(2k) \ln\left( {\frac{{2k_F+2k+i\delta}}{{2k+i\delta}}}\right)\right.}&\cr &
\displaystyle{\left.-r_0^*V(2k) \ln\left( {\frac{{2k-2k_F+i\delta}}{{
2k+i\delta}}}\right)\right),}&  \nonumber
\end{eqnarray}
and 
\begin{eqnarray}
&\displaystyle{\delta r_a^{(2)} = -{\frac{{mr_a r_0^*}}{{4\pi k^*}}} \left(
r_0 V(-2k^*) \ln\left( {\frac{{2k_F-2k^*+i\delta}}{{-2k^*+i\delta}}}
\right)\right.}&  \nonumber \\
&\displaystyle{\left.-r_0^*V(-2k^*) \ln\left( {\frac{{-2k^*-2k_F+i\delta}}{{
-2k^*+i\delta}}}\right)\right).}&  \nonumber
\end{eqnarray}
Keeping only the divergent terms, we are finally left with 
\begin{equation}
\delta r_a = {\frac{{r_a r_0 r_0^* }}{{2\pi v}}} V(2k_F) \ln\left({\frac{{E_F}
}{{\epsilon}}}\right).
\end{equation}
As it is usual in 1D systems, perturbative corrections are logarithmically
divergent, signalling an instability of the Fermi liquid ground state and
the emergence of the Luttinger liquid behavior. One can then sum the leading
logarithmic singularity by deriving and solving the renormalization group
equation. Here we derive an equation for the Andreev reflection coefficient 
$R_a =|r_a|^2$. The subgap conductance is obtained via the formula $G = 4\pi
e^2 R_a$, where the condition $R_a + R_0 =1$ has been used. By defining 
$\alpha = {\frac{{V(2k_F )}}{{2\pi v}}}$ one obtains the equation 
\begin{equation}
{\frac{{d R_a }}{{d\eta }}} = 2 R_a (1- R_0 ) ,  \label{rg}
\end{equation}
where $\eta =\ln{\Delta /\epsilon }$ with initial condition $\eta =0$ 
($\epsilon = \Delta $). The solution of Eq.~(\ref{rg}) gives 
\begin{equation}
R_a(\epsilon )=\displaystyle{\ {\frac{{R_a (\Delta /\epsilon )^{2\alpha }}}
{{R_0 +R_a (\Delta / \epsilon )^{2\alpha}}}} }.  \label{solution}
\end{equation}
Here $R_a$ and $R_0$ are the initial values of the scattering coefficients.
Our analysis has neglected the cutoff dependence of the interaction
coupling, which a more complete calculation should take into account, as
well as the Fock terms. In Ref.~\cite{takane97} this calculation has been
carried out and the result (\ref{solution}) remains unchanged provided 
$\alpha \rightarrow \alpha -2 \beta$ with $\beta = {\frac{{V(0)}}{{2\pi v}}}$.

\appendix

\section{One-particle correlation function}

\label{onepart}

In this Appendix, we present some explicit expressions for the integrated
single particle Green's function of an infinitely long LL, defined as 
$
G(x) = \int \limits _{-\beta /2} ^{\beta/2} d\tau G_{s,s}(x,\tau) , 
$
where $G_{s,s^{\prime}}(x,\tau) \equiv - \langle T_{\tau} \psi _{s}(x,\tau)
\psi ^{\dagger}_{s^{\prime}}(0,0) \rangle . $ The latter quantitity is
readily evaluated~\cite{voit95}, using the Bose representation~(\ref
{fieldoperator}). Upon integration over $\tau$ one obtains in the long
wavelength limit 
\begin{equation}
G(x) = \frac{2}{v_{F}} \sin (k_{F}x) \left(\frac{\alpha}{x}
\right)^{1/2g_{\rho}+g_{\rho}/8 - 1/2} F_{G}(x,g_{\rho}).  \label{Gint}
\end{equation}
In the zero temperature limit, $F_{G}$ is $x$-independent, 
\begin{eqnarray}
&&F_{G}(x,g_{\rho}) \to \int \limits _{-\infty}^{\infty} \frac{dz}{\pi} \sin
\left[\frac{\arctan (1/z) + \arctan (g_{\rho}/2z)}{2}\right]  \nonumber \\
&&\times \left[ \frac{1}{1+z^{2}}\right]^{1/4} \left[ \frac{1}{
1+(2z/g_{\rho})^{2}}\right]^{1/4g_{\rho} + g_{\rho}/16} \equiv F_G(g_\rho).
\label{FGzero}
\end{eqnarray}
Note that $F_{G}(2) =1$. For noninteracting electrons, $F_{G}(x,g_{\rho})$
can be found explicitly, 
\begin{equation}
F_{G}(x,2) = \frac{2}{\pi} \arctan[1/\sinh(\pi x/\beta v_{F})].
\label{FGnonint}
\end{equation}
In the case of weak repulsive interactions, $(2-g_{\rho})\pi x/2\beta v_{F}
\ll 1$ and low temperatures $\pi x/2\beta v_{F} \ll 1$, the function 
$F_{G}(x,g_{\rho})$ can be approximated as 
\begin{equation}
F_{G}(x,g_{\rho}) \approx F_{G}(g_{\rho}) \left[ 1 - \frac{2x}{\beta v_{F}}
+ \frac{2-g_{\rho}}{2} \frac{\pi x}{\beta v_{F}} \right] .  \label{FGweak}
\end{equation}
Here we dropped terms ${\cal O} [(2-g_{\rho})(\pi x/\beta v_{F})^{2} +
(2-g_{\rho})^{2}]$.

\section{Two-particle correlation function}

\label{twopart}

Next we consider the two-particle correlation function $\Pi _{C} (x,\tau)
\equiv \Pi _{+,-,-,+}(x,\tau ;x,\tau; 0,0;0,0) , $ which describes
particle-particle propagation of a spin singlet in a LL over a distance $x$
during a time $\tau $. Integrated over imaginary time, this quantitiy can be
written as 
\[
\Pi _{C} (x) = \int \limits _{-\beta/2} ^{\beta/2} d\tau \Pi _{C} (x,\tau) =
\rho _{0} \frac{2}{v_{F}} \left( \frac{\alpha}{x} \right)^{2/g_{\rho}}
F_{C}(x,g_{\rho}) . 
\]
At zero temperature we find $F_{C}(x,g_{\rho}) \to F_C(g_\rho)$, with 
\begin{equation}
F_{C}(g_{\rho}) = \int \limits _{-\infty} ^{\infty} \frac{dz}{\pi} \left[ 
\frac{1}{1+z^{2}} \right] ^{1/2} \left[ \frac{1}{1+(2z/g_{\rho})^{2}}
\right]^{1/g_{\rho}} .  \label{FCzero}
\end{equation}
Note that $F_{C}(2) =1$. In the noninteracting case $F_{C}(x,g_{\rho})$ can
be calculated analytically also at finite temperatures, 
\begin{equation}
F_{C}(x,2) = \frac{(2\pi x/\beta v_{F})}{\sqrt{\cosh ^{2}(2\pi x/\beta
v_{F}) -1}} .  \label{FCnonint}
\end{equation}
In the limit of high temperatures, $2\pi x/(\beta v_{F}) \gg 1$, $F_{C}$ can
be found for arbitrary interaction strength, 
\begin{equation}
F_{C}(x,g_{\rho)}) \approx 2 \left(\frac{2\pi x}{\beta v_{\rho}}
\right)^{2/g_{\rho}} \exp{(-2\pi x/\beta v_{F})} .  \label{FhiT}
\end{equation}
In the case of weak repulsive interactions and low temperatures an expansion
similar to the one leading to (\ref{FGweak}) enables us to approximate 
\begin{equation}
\frac{F_{C}(x,g_{\rho})}{F_{C}(g_{\rho})} \simeq 1 - \frac{2}{3} \left(\frac{
\pi x}{\beta v_{F}}\right)^{2} + (2-g_{\rho}) \left(\frac{\pi x}{\beta v_{F}}
\right)^{2/g_{\rho}}.  \label{FCweak}
\end{equation}

\end{multicols}


\begin{references}
\bibitem{intnote}  It is assumed that the Coulomb potential is screened,
e.g., by a back-gate, on distances larger than $\lambda _{F}$, but shorter
than the long wavelengths of interest here.
\end{references}
\end{document}